\documentclass[twoside,fleqn]{article}
\usepackage{npb,epsfig}

\newcommand{\AmS}{{\protect\the\textfont2
  A\kern-.1667em\lower.5ex\hbox{M}\kern-.125emS}}

\title{N$^3$LO fits to $xF_3$ data: $\alpha_s$ 
vs $1/Q^2$ contributions
}

\author{A. L. Kataev \address{Institute for Nuclear Research of the Academy 
of Sciences of Russia,
         \\
        11312, Moscow, Russia}
        \thanks{Supported by RFBI Grants N 00-02-17432, 02-01-00601
               }, 
        G. Parente \address{Department of Particle Physics, University 
        of Santiago de Compostela, \\ 15706 Santiago de Compostela, Spain}
        \thanks{Supported by Xunta de Galicia (PGIDT00PX20615PR) and 
         CICYT (AEN99-0589-C02-02)} 
        and 
        A. V. Sidorov \address{Bogoliubov Laboratory of Theoretical Physics,
        Joint Institute for Nuclear Research, \\
        141980 Dubna, Russia}\thanks{Supported by RFBI Grants
N 00-02-17432, 02-01-00601  and INTAS (Call 2000-project N 587)}}

\begin{document}

\begin{abstract}
The results of approximate N$^3$LO and detailed NNLO fits to 
$xF_3$ data of the  CCFR'97 collaboration are presented. 
We demonstrate that $1/Q^2$ non-perturbative corrections 
to $xF_3$ modeled by three independent procedures 
are shadowed by perturbative QCD effects, starting at  the NNLO. 
Special attention 
is paid to revealing the role  
of the recently calculated NNLO corrections to the anomalous dimensions 
and N$^3$LO corrections to the coefficient functions of odd moments 
of $xF_3$ with $n\leq 13$.
The related values of $\alpha_s(M_Z)$ are extracted. 
\end{abstract}

\maketitle

It is known that the leading non-perturbative power suppressed corrections 
to DIS structure functions (SFs) have the dimension $1/Q^2$. 
However, it turned out that phenomenological value of non-perturbative 
effects depend crucially from the order of the corresponding perturbative 
contributions. It is worth to remind that in 1979, when the data for DIS 
neutrino-nucleon scattering was not precise enough, the authors of Ref. 
\cite{Abbott:1979as} 
were unable to separate perturbative $1/\ln(Q^2)$ source of scaling violation 
from  the $1/Q^2$-effects. At present both the precision of  
 $xF_3$ measurements 
\cite{Seligman:mc} and the information on 
renormalization-group 
perturbative QCD evolution of Mellin 
moments became more precise. 
The latter ones enter into  the Jacobi polynomial 
formula \cite{Parisi:1978jv}
\begin{eqnarray}
\label{Jacobi}
xF_3(x,Q^2)&=&w(\alpha,\beta)\sum_{n=0}^{N_{max}}\Theta_n^{(\alpha,\beta)}(x)
\times \\  
&&\sum_{j=0}^{n}c_j^{(n)}(\alpha,\beta)M_{j+2}^{TMC}(Q^2)+\frac{HT}{Q^2}
\nonumber 
\end{eqnarray} 
where $w(\alpha,\beta)=x^{\alpha}(1-x)^{\beta}$,  
$c_j^{(n)}(\alpha,\beta)$ contain Euler 
$\Gamma$ functions from $\alpha$ and $\beta$ and the moments 
\begin{equation}
M_n^{TMC}(Q^2)=M_n(Q^2)+\frac{n(n+1)M_{nucl}^2}{(n+2)Q^2}
\end{equation}
 are taking into account the leading order target mass corrections. 
The kinematic contributions of order $1/Q^4$ did not affect 
the results 
of our previous less detailed NNLO fits to CCFR'97 data  
(see Refs.\cite{Kataev:1997nc,Kataev:1999bp})
and the analysis of Ref.\cite{Kataev:2001kk}
described below.

In this talk we will concentrate  
on the results of the most recent analysis 
of the CCFR'97 data for $xF_3$ in the NLO, NNLO and approximate N$^3$LO levels 
of perturbative QCD \cite{Kataev:2001kk}, paying special attention to the 
possibility of the detection of non-perturbative $1/Q^2$-contributions 
to $xF_3$. They will be  modeled 
by three independent ways. First, is the infrared-renormalon (IRR) 
model  of Ref. \cite{Dasgupta:1996hh}
\begin{equation}
\frac{HT}{Q^2}=w(\alpha,\beta)\sum_{n=0}^{N_{max}}
\Theta_n^{(\alpha,\beta)}\sum_{j=0}^nc_j^{(n)}(\alpha,\beta)M_{j+2}^{IRR}
\end{equation}
where  $M_n^{IRR} =\tilde{C}M_n(Q^2)A_2^{'}/Q^2$ 
and  $\tilde{C}=-n+4+2/(n+1)+4(n+2)+4S_1(n)$,  calculated in 
Ref. \cite{Dasgupta:1996hh} from the single chain of quark loop insertions 
into the one-gluon contribution to the corresponding Born diagram and 
$A_2^{'}$ is the arbitrary fitted parameter. 
Next, following  
the NNLO Bernstein polynomial fits to CCFR'97 data of Ref. 
\cite{Santiago:2001mh} (see also Ref. \cite{Maxwell:2002mt}) 
we  consider gradient model of twist-4 term,
namely 
\begin{equation}
M_{n,xF_3}^{HT}(Q^2)=n\frac{B_2^{'}}{Q^2}M_n(Q^2)
\end{equation}
where $B_2^{'}$  is the free parameter. 
Another possibility is 
to choose  
\begin{equation}
HT=h(x) {\it~ in~ the~ model-independent~ way}~.
\end{equation}
Here $h(x)$  is defined by free parameters $h_i=h(x_i)$, where $x_i$ are the 
points in experimental data binning. 
In our work the following renormalization-group equation 
for 
the Mellin 
moments of $xF_3$ was used:  
\begin{equation}
\frac{M_n(Q^2)}{M_n(Q_0^2)}=exp\bigg[-\int\frac
{\gamma_{F_3}^{(n)}(t)}{\beta(t)}dt\bigg]\frac{C_{F_3}^{(n)}(A_s(Q^2))}
{C_{F_3}^{(n)}(A_s(Q_0^2))}
\end{equation}   
where $A_s=\alpha_s/(4\pi)$ is the $\overline{MS}$-scheme coupling 
constant and $M_n(Q_0^2)$ is defined in the initial scale 
as $M_n(Q_0^2)=\int_0^1x^{n-2}A(Q_0^2)x^{b(Q_0^2)}(1+\gamma(Q_0^2))dx$

At the N$^3$LO the expression for $C_{F_3}^{(n)}(A_s)$ can be presented 
in the following form
\begin{equation}
\label{C3}
C_{F_3}^{(n)}=1+C^{(1)}(n)A_s+C^{(2)}(n)A_s^2+C^{(3)}(n)A_s^3
\end{equation}
where the NNLO correction  $C^{(2)}(n)$ can be obtained for any $n$ 
from the results of Ref. \cite{Zijlstra:1992kj}, which were 
checked with the help of other methods in Ref. \cite{Moch:1999eb}.
The N$^3$LO contributions to Eq. (\ref{C3}) were analytically 
calculated in Ref. \cite{Retey:2000nq} for odd $n\leq 13$.
The  N$^3$LO expansion of the anomalous dimension term
has the following form  
\begin{eqnarray}
\nonumber
&&exp\bigg[\int^{A_s(Q^2)}\frac{\gamma_{F_3}^{(n)}(t)}{\beta(t)} dt\bigg]=
\\ \nonumber 
&&=(A_s(Q^2))^{\gamma_{F_3}^{(0)}/\beta_0}\times \bigg[1+p(n)A_s(Q^2)
\\ \label{qn} 
&&+q(n)A_s(Q^2)^2+r(n)A_s(Q^2)^3\bigg]
\end{eqnarray}
where $p(n)$, $q(n)$ and $r(n)$ are defined through coefficients 
of QCD $\beta$-function 
 and anomalous dimension 
$\gamma_{F_3}^{(n)}$ (see Ref. \cite{Kataev:2001kk}). Note, that on the 
contrary to the QCD $\beta$-function, analytically calculated 
in Ref. \cite{vanRitbergen:1997va} at the N$^3$LO 
level, the expression for 
$\gamma_{F_3}^{(n)}$ is known up to NNLO order. Moreover, 
its NNLO corrections were calculated in case of odd $n\leq 13$ only. 
In order to fix the numerical values of the 
 NNLO corrections to $\gamma_{F_3}^{(n)}$
(and thus the term $q(n)$)  for even $n$ inside the interval 
 $3\leq n\leq 13$ we used the  smooth interpolation procedure, proposed 
in Ref. \cite{Parente:1994bf}, and supplemented it  by fine-tuning of 
definite NNLO coefficients of $\gamma_{F_3}^{(n)}$ \cite{Kataev:2001kk} . 
The application 
of this procedure for estimating NNLO coefficients of $\gamma_{F_3}^{(n)}$ 
with $n$  even and $N_f=4$  result in the numbers, which differ from the  
corresponding NNLO terms of non-singlet contributions to 
$\gamma_{F_2}^{(n)}$ \cite{Larin:1996wd,Retey:2000nq} in the 4th significant 
digit. 
The NNLO correction to $\gamma_{F_3}^{(2)}$ was 
estimated by us using extrapolation procedure, which has definite theoretical 
uncertainties.
As to the applicability of  the smooth 
interpolation procedure, 
we checked 
that it  is reproducing 
the known even contributions to $C^{(2)}(n)$ with satisfactory precision 
\cite{Kataev:2001kk}. That is why we consider the results of its 
application, including the estimates of  even terms of  
$C^{(3)}(n)$,  as really reliable.

Fixing by this way  the N$^3$LO coefficients $C^{(3)}(n)$ 
and  estimating N$^3$LO correction 
$r(n)$ to Eq. (8) by means of [1/1] Pad\'e approximation technique, 
previously used in perturbative QCD e.g. in Ref. \cite{Brodsky:1997vq}, 
we can use Eq. (1) for performing approximate N$^3$LO fits 
to $xF_3$ data. At the next page the results, obtained in 
Ref. \cite{Kataev:2001kk} in 
the case of combining Eq. (1) with the IRR model of Eq. (3), are presented  
for  $N_{max}=6$, first studied in 
Refs. \cite{Kataev:1997nc,Kataev:1999bp}, and $N_{max}=9$ (see Table 1).  

Looking at Table 1 we arrive at the following conclusions:

1) The NLO fits seem to support the IRR model of Ref. \cite{Dasgupta:1996hh} by 
the foundation of the negative values of $A_2^{'}$. 

\begin{table*}[hbt]
\setlength{\tabcolsep}{1.5pc}
\newlength{\digitwidth} \settowidth{\digitwidth}{\rm 0}
\caption{The results to the fits of CCFR'97 data.  
$A_2^{'}$ $[{\rm GeV^2}]$ is the IRR model parameter.
$\Lambda_{\overline{MS}}^{(4)}$ is measured in ${\rm MeV}$.
The cases of different $Q_0^2$ and $N_{max}$ are considered.}
\catcode`?=\active \def?{\kern\digitwidth}
\label{tab1}
\begin{tabular*}{\textwidth}{@{}l@{\extracolsep{\fill}}rrrr}
\hline
 {order/$N_{max}$} & $Q_0^2=$ & $5~{\rm GeV}^2$ & $20~{\rm GeV}^2$ & 
$100~{\rm GeV}^2$ \\                   
\hline
 NLO/6    & $\Lambda_{\overline{MS}}^{(4)}$ & 370$\pm$38  & 369$\pm$41  
& 367$\pm$38 \\
         & $\chi^2/nep$ & 80.2/86 & 80.4/86        & 79.9/86 \\
     & $A_2^{'}$ & $-$0.121$\pm$0.052 & $-$0.121$\pm$0.053 
& $-$0.120$\pm$0.052 \\
 NLO/9   & $\Lambda_{\overline{MS}}^{(4)}$ & 379$\pm$41  & 376$\pm$39  
& 374$\pm$42 \\
         & $\chi^2/nep$ & 78.6/86 & 79.5/86        & 79.0/86 \\
     & $A_2^{'}$ & $-$0.125$\pm$0.053 & $-$0.125$\pm$0.053 
& $-$0.124$\pm$0.053 \\
\hline
 NNLO/6    & $\Lambda_{\overline{MS}}^{(4)}$ & 297$\pm$30  & 328$\pm$36  
& 328$\pm$35 \\
         & $\chi^2/nep$ & 77.9/86 & 76.8/86        & 79.5/86 \\
     & $A_2^{'}$ & $-$0.007$\pm$0.051 & $-$0.017$\pm$0.051 
& $-$0.015$\pm$0.053 \\
 NNLO/9   & $\Lambda_{\overline{MS}}^{(4)}$ & 331$\pm$33  & 332$\pm$35  
& 331$\pm$35 \\
         & $\chi^2/nep$ & 73.1/86 & 75.7/86        & 76.9/86 \\
     & $A_2^{'}$ & $-$0.013$\pm$0.051 & $-$0.015$\pm$0.051 
& $-$0.016$\pm$0.051 \\
\hline
 N$^3$LO/6    & $\Lambda_{\overline{MS}}^{(4)}$ & 305$\pm$29  & 327$\pm$34  
& 326$\pm$34 \\
         & $\chi^2/nep$ & 76.0/86 & 76.2/86        & 78.5/86 \\
     & $A_2^{'}$ & 0.036$\pm$0.051 & 0.033$\pm$0.052 
& 0.029$\pm$0.052 \\
 N$^3$LO/9   & $\Lambda_{\overline{MS}}^{(4)}$ & 333$\pm$34  & 328$\pm$33  
& 328$\pm$38 \\
         & $\chi^2/nep$ & 73.8/86 & 75.9/86        & 76.4/86 \\
     & $A_2^{'}$ & 0.038$\pm$0.052 & 0.035$\pm$0.052 
& 0.034$\pm$0.052 \\
\hline
\end{tabular*}
\end{table*}  
These  values are  in agreement with 
the results of the previous fits of Refs. \cite{Kataev:1997nc,Kataev:1999bp}
and with the ones, obtained in Ref. \cite{Alekhin:1998df} using the NLO 
DGLAP analysis of the same set of CCFR'97 data and the parton distributions
set (PDFs) of Ref. \cite{Alekhin:1996za}. The similar value 
$A_2^{'}=-0.104\pm 0.005~{\rm GeV^2}$ was also found in the NLO 
fits to the combined $F_2$ data \cite{Yang:1999xg} using MRS(R2) PDFs     
\cite{Martin:1996as}.

2) At the NNLO the values of $A_2^{'}$ are comparable wit zero within 
statistical error bars. The similar small value, 
namely $A_2^{'}=-0.0065\pm0.0059$,  was obtained from the NNLO fits 
to $F_2$ data in Ref. \cite{Yang:1999xg}.

3) The inclusion of the N$^3$LO corrections make $A_2^{'}$ positive. However,
it  has 
the statistical uncertainties twice as large as the central value.

Thus we conclude that starting from the NNLO the IRR-model corrections
to $xF_3$ can not be extracted from CCFR'97 data with reasonable 
precision and are shadowed by perturbative QCD effects. 

4) The values of $\chi^2$ decrease from NLO up to NNLO and at the 
N$^3$LO it almost coincide with the ones obtained at the NNLO. Moreover, 
$\chi^2$ decreases with the increase of $N_{max}$. This is the welcome 
feature of including in the fits more detailed information on the perturbative
theory contributions both to the coefficient functions and the anomalous 
dimensions of $xF_3$ moments.       

5) For $N_{max}=9$ $\Lambda_{\overline{MS}}^{(4)}$ and $A_2^{'}$ are rather 
stable to variation of $Q_0^2$ not only at the NLO but at the NNLO and 
N$^3$LO as well. This property gives favor of our new results from 
Ref. \cite{Kataev:2001kk} in comparison with the ones obtained in  
Ref. \cite{Kataev:1999bp} for $N_{max}=6$ and $Q_0^2=20~{\rm GeV^2}$
using more approximate model for $\gamma_{F_3}^{(2)}(n)$ and Pad\'e 
approximation for $C_{F_3}^{(3)}(n)$.

To transform $\Lambda_{\overline{MS}}^{(4)}$ into the values of 
$\alpha_s(M_Z)$ we first used the $\overline{MS}$-scheme matching condition 
of Ref. \cite{Chetyrkin:1997sg} with the matching point chosen as 
$m_b^2\leq M_b^2\leq 36m_b^2$ following the proposal of 
Ref. \cite{Blumlein:1998sh}. This gives us the possibility to estimate 
threshold uncertainties in $\alpha_s(M_Z)$. 
Varying the factorization and renormalization scales 
$\mu^2_R=\mu^2_{FAC}=\mu^2$ in the interval $\mu_{\overline{MS}}^2/4\leq
\mu_{\overline{MS}}^2\leq 4 \mu_{\overline{MS}}^2$ we estimated the 
scale-dependent uncertainties. 
As the result we obtained 
the following values of $\alpha_s(M_Z)$ extracted from the fits to $xF_3$ 
CCFR'97 data with $1/Q^2$ non-perturbative corrections modeled using 
the IRR approach \cite{Kataev:2001kk}:  
\begin{eqnarray}
\nonumber
&&NLO~~\alpha_s(M_Z)=0.120\pm 0.002~(stat)\\ \nonumber
&& \pm 0.005~(syst) 
  \pm0.002~(thresh)^{+0.010}_{-0.006}~(scale) \\  
\label{alpha} 
&&NNLO~~\alpha_s(M_Z)=0.119\pm 0.002~(stat) \\ \nonumber
&& \pm 0.005~(syst) 
\pm0.002~(thresh)^{+0.004}_{-0.002}~(scale) \\ \nonumber 
&&N^3LO~~\alpha_s(M_Z)=0.119\pm 0.002~(stat) \\ \nonumber  
&&\pm 0.005~(syst) 
\pm0.002~(thresh)^{+0.002}_{-0.001}~(scale) 
\end{eqnarray}
The systematic errors are fixed from separate consideration of these 
experimental uncertainties of the  CCFR'97 collaboration. Notice, that the 
inclusion of higher-order perturbative QCD corrections into  Eq. (6) minimize 
essentially the scale-dependence uncertainties. They are in agreement 
with the similar estimates, obtained in Ref. \cite{vanNeerven:2001pe}
in the process of the fits to the definite model of $xF_3$ data.
Our NNLO value of Eq. (10), obtained in Ref. \cite{Kataev:2001kk},
within existing error-bars is in agreement with the results of NNLO 
Bernstein polynomial fits to the CCFR'97 data, namely 
$\alpha_s(M_Z)=0.1153\pm 0.0041~(exp)\pm 0.0061 ~(theor)$ 
\cite{Santiago:2001mh} and $\alpha_s(M_Z)=0.1196^{+0.0027}_{-0.0031}$
\cite{Maxwell:2002mt}.

In Table 2 we present the outcomes of our NLO and NNLO  
fits to the subset of  CCFR'97 data, analyzed in Ref.\cite{Santiago:2001mh}. 
The  twist-4 term was  fixed with the 
help of the gradient model of Eq. (4).  
\\ [-10mm]
\begin{table}[hbt]
\catcode`?=\active \def?{\kern\digitwidth}
\caption{
 The results of the fits to the subset of  CCFR'97 data with 
HT defined by 
the gradient  model with the coefficient $B_2^{'}$ [ $\rm {GeV}^2$].}
\label{tab:effluents1}
\begin{tabular}{lrcr}
\hline
                ${\rm order}$           &
                $\Lambda_{\overline{MS}}^{(4)}$ (MeV)     &
                $ B_2^\prime$(HT)   &
 $\chi^2/nep$  \\
\hline
NLO   & 371$\pm$72      & -0.135$\pm$0.113 &
75/74     \\
NNLO   & 316$\pm$51      & -0.031$\pm$0.088 &  
64/74\\
\hline
\end{tabular}
\end{table}
\\[-10mm]
At the NLO the value of $B_2^{'}$ is in agreement with the 
value of the IRR model parameter $A_2^{'}$. At the NNLO $B_2^{'}$ is 
comparable  with zero. Thus we confirm the existence of the 
effect of the shadowing of the 
dynamical $1/Q^2$-corrections to $xF_3$ SF  at the  NNLO 
of perturbative QCD. This effect  
was first observed using the IRR model in  Ref. \cite{Kataev:1997nc}.  

If we  use model-independent 
parameterization of the twist-4 contribution (see Eq. (5)), 
this effect is becoming even more vivid. The results 
of extraction of $h(x)$ in different orders of perturbative QCD 
and for different $N_{max}$  are presented at 
Fig.1, taken from Ref. \cite{Kataev:2001kk}.    
\begin{figure}[hbt]
\vspace{-20pt}
\psfig{width=7cm,file=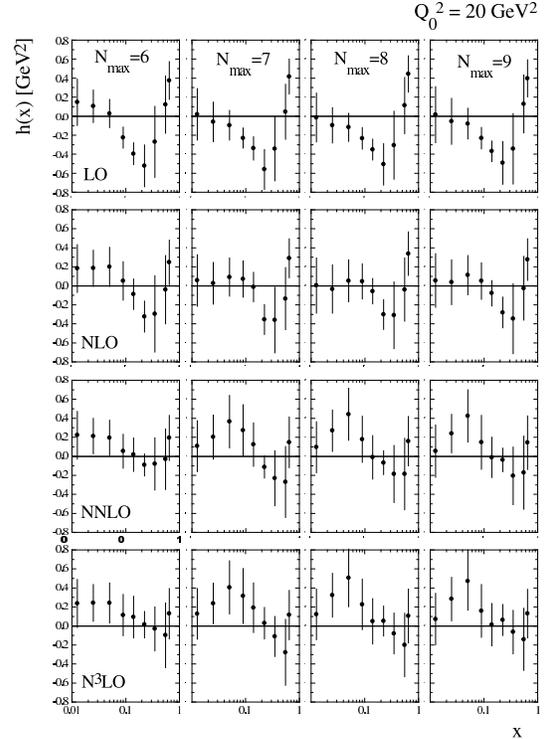}
\caption{The $x$-shape of $h(x)$ extracted from the
fits of CCFR'97 for $Q_0^2=20~GeV^2$.}
\label{fig:largenenough}
\end{figure}
\\[-10mm]
Several comments are in order.

1) The $x$-shape of $h(x)$, obtained at the LO and NLO, is in satisfactory 
agreement with the prediction of the IRR model of Ref. \cite{Dasgupta:1996hh}.

2) In all orders of perturbation theory the results are  rather stable to 
the variation of $N_{max}$.

3) The $x$-shape of $h(x)$, obtained during the NNLO and approximate 
N$^3$LO fits, demonstrate oscillation-type behavior 
with large error-bars. Thus we conclude, that starting from the 
NNLO the $x$-shape of $h(x)$ is 
strongly correlated with                            
higher-order perturbative 
QCD corrections to Eq. (6).

However, it is possible that more detailed understanding of the 
NNLO behavior of $h(x)/Q^2$ contribution to $xF_3$  
will   be obtained after NNLO  analysis with taking 
into account systematic uncertainties of the data \cite{Alekhin}.
 
To conclude, we demonstrated that the inclusion into the fits 
to CCFR'97  $xF_3$ data of the  NNLO perturbative QCD effects is 
leading to effective shadowing 
of the  dynamical $1/Q^2$-corrections. However, it might be possible, 
that they 
will  be detected in future even at the NNLO. 
This might happen in case  more precise experimental data 
for the SFs of $\nu N$ DIS will be obtained, say   
at the future neutrino factories (for detailed discussions  
see Ref. \cite{Mangano:2001mj}).
 
Note also, that 
while  considering  massive-dependent  
perturbative series for the      Adler function 
of  $e^+e^-$-scattering  \cite{Eidelman:1998vc}  and 
re-extracting   the gluon condensate value from the  charmonium sum rules 
\cite{Ioffe:2002be}
with the calculated in Ref. \cite{Chetyrkin:1996cf}
three-loop massive corrections  to their spectral function 
the effects of influence of  higher-order perturbative QCD effects 
to  the value of the  gluon condensate, introduced in Ref. \cite{Shifman:bx},
were observed.
Thus, the  problems of correlations between 
perturbative and non-perturbative terms, discussed in our talk,  
seem to be typical to other cases also, 
when it is necessary to analise the expansions in both $\alpha_s$ and inverse 
powers of $Q$. The similar point of view was expressed at this Symposium 
in the talk of Ref. \cite{Glover}.

{\bf Acknowledgments}
One of us (ALK) is grateful to the members of OC of  RADCOR-02
and Loops-and-Legs in Quantum Field Theory Symposium  
for hospitality and financial support.
We are grateful to S.I. Alekhin and A.A. Pivovarov for discussions.

\end{document}